\documentclass[aps,twocolumn,pra,superscript,floatfix,superscriptaddress,showpacs,footinbib,English]{revtex4-1}
\usepackage{amssymb}

\DeclareUnicodeCharacter{0308}{\"{}}
\UseRawInputEncoding

\usepackage[pdftex]{graphicx}
\usepackage{dcolumn}% Align table columns on decimal point
\usepackage{bm}% bold math
\usepackage{amsmath}
\usepackage{array}
\usepackage{color}
\usepackage{float}
\usepackage{subfigure}
\usepackage{dsfont}
\usepackage{txfonts}
\usepackage{wasysym}
\usepackage{multirow}
\usepackage{sidecap}
\usepackage{xcolor,cancel}
\usepackage[normalem]{ulem}

\newcommand{\bpng}{BPN/G}
\newcommand{\bpngn}{BPN/G$_\text{n}$}
\newcommand{\bpngh}{(BPN/G$_\text{n}$)$_\text{4H}$}

\usepackage{hyperref}
\hypersetup{
    colorlinks,%
    citecolor=blue,%
    linkcolor=blue,%
    urlcolor=blue
}

\begin{document} 

\title{Catenary-like rippled biphenylene/graphene lateral heterojunction}

%\title{First-principles insights into curved graphene/biphenylene lateral junctions}

\author{Victor M. S. da C. Dias}
\affiliation{Instituto de F\'{\i}sica,  Universidade Federal de Uberlândia, Av. João Naves de Ávila, 2121 - MG, 38400-902, Brazil}

\author{Danilo de P. Kuritza}
\affiliation{Campus Avançado Jandaia do Sul, Universidade Federal do Paraná,
86900-000, Jandaia do Sul, PR, Brazil}

\author{Igor S. S. de Oliveira}
%\email{igor.oliveira@ufla.br}
\affiliation{Departamento de F\'isica, Universidade Federal de Lavras, C.P. 3037, 37203-202, Lavras, MG, Brazil}

\author{Jose E. Padilha}
\affiliation{Campus Avançado Jandaia do Sul, Universidade Federal do Paraná,
86900-000, Jandaia do Sul, PR, Brazil}

\author{Roberto H. Miwa}
\email{hiroki@ufu.br}
\affiliation{Instituto de F\'{\i}sica,  Universidade Federal de Uberlândia, Av. João Naves de Ávila, 2121 - MG, 38400-902, Brazil}

%\author{\textcolor{red}{Who else?}}

\date{\today}

\begin{abstract}

In this study, we conduct a first-principles analysis to explore the structural and electronic properties of curved biphenylene/graphene lateral junctions (\bpng). We start our investigation focusing on the energetic stability of \bpng\, by varying the width of the graphene region, \bpngn. The electronic structure of \bpngn\, reveals (i) the formation of metallic channels mostly localized along the BPN stripes, where (ii) the features of the energy bands near the Fermi level are ruled by the width (n) of the graphene regions, G$_\text{n}$. In the sequence, we find that the hydrogenation of \bpngn\, results in a semiconductor system with a catenary-like rippled geometry. The electronic states of the hydrogenated system are mainly confined in the curved G$_\text{n}$ regions, and the dependence of the bandgap on the width of G$_\text{n}$ is similar to that of hydrogenated armchair graphene nanoribbons. The effects of curvature on the electronic structure, analyzed in terms of external mechanical strain, revealed that the increase/decrease of the band gap is also dictated by the width of the G$_\text{n}$ region. Further electronic transport calculations reveal a combination of strong transmission anisotropy and the emergence of negative differential resistance. Based on these findings, we believe that rippled biphenylene/graphene systems can be useful for the design of two-dimensional nanodevices.

\end{abstract}
\maketitle
\section{Introduction}

Graphene, a monolayer of sp$^2$-hybridized carbon atoms arranged in a two-dimensional (2D) honeycomb lattice, has attracted immense interest in the scientific and technological fields over the last two decades due to its exceptional properties. These include remarkable electrical conductivity, mechanical strength, flexibility, and thermal conductivity\,\cite{novoselov2004, geim2, Novoselov2012, grapheneconductive, Balandin2008}, which have established graphene as a material with immense potential for use in a wide range of advanced technologies\,\cite{tiwari2020graphene}. 

Beyond graphene, numerous 2D carbon-based allotropes have been proposed, each offering distinct electronic and mechanical properties. Among them, graphyne and graphdiyne are noteworthy for their acetylenic linkages, which introduce unique characteristics\,\cite{kang2018graphyne, li2014graphdiyne}. Other forms of graphene allotropes\,\cite{enyashin2011graphene, jana2021emerging}, such as biphenylene, have recently emerged as intriguing materials for both fundamental research and practical applications. While graphene exhibits semi-metallic properties due to its unique band structure, biphenylene is a metallic material, formed by a periodic arrangement of octagons, hexagons, and squares with sp$^2$ hybridization, which guarantees robust mechanical properties\,\cite{Fan2021, al2022two, liu2021two, liu2021type, han2022biphenylene, luo2021first}.

For certain applications, introducing a band gap into these materials becomes essential. For instance, applications in transistors, optoelectronics (such as photodetectors and light-emitting diodes), solar cells, and integrated circuits require a band gap. Various methods have been developed to open a band gap in graphene, including strain engineering, chemical functionalization, hydrogenation, graphene nanoribbons, stacking with other 2D materials, and applying an electric field perpendicular to the graphene plane\,\cite{nandee2022band, han2007energy, pereira2009strain, nandee2022band, lu2010band,gao2011band}. The band gap opening in biphenylene has also been the subject of recent studies, showing that it can be achieved through strain engineering\,\cite{Hou2023} or hydrogenation\,\cite{lee2021band}.

Hydrogenation, in particular, has proven to be a versatile technique for modifying the properties of graphene. The transition from planar sp$^2$ to sp$^3$ hybridization upon hydrogen adsorption induces significant changes in the structural and electronic properties, making graphene semiconducting when fully hydrogenated\,\cite{boukhvalov2008hydrogen, whitener2018hydrogenated, cadelano2010elastic, lebegue2009accurate}. This opens up possibilities for tailored functionalities in graphene-based technologies.

The integration of 2D materials into lateral heterojunctions has opened new avenues for band structure engineering, enabling the development of advanced electronic devices\,\cite{dutta2024electronic, haidari2024graphene, kim2020spatially}. Unlike vertical heterostructures, lateral junctions offer reduced capacitance and an expanded space charge region, providing unique advantages for device design\,\cite{li2024edge}. Due to the distinct electronic properties found in graphene and biphenylene, the combination of these two materials forms an ideal platform for exploring heterojunctions with tunable electronic properties. 
%The introduction of a bandgap is often essential for applications in transistors, optoelectronic devices, solar cells, and integrated circuits. Various methods, including strain engineering, chemical functionalization, hydrogenation, and nanostructuring, have been developed to achieve bandgap opening in graphene and related materials\,\cite{nandee2022band, han2007energy, pereira2009strain, lu2010band, gao2011band}. 

Furthermore, the unique transport properties of graphene and other 2D materials have been the focus of significant research. These properties can be tuned by external factors such as curvature, strain, and heterostructure formation, paving the way for advanced applications in transistors, sensors, and quantum information technologies\,\cite{sangwan2018electronic, nourbakhsh2016transport, qi2023recent}.

The role of curvature in graphene has also been extensively studied, revealing its profound impact on physical properties and potential applications. Curvature engineering has been leveraged to create strain-sensitive devices, enhance catalytic activity, and design stretchable electronics\,\cite{huang2018assembly, wang2011super, liu2020substrate, ortolani2012folded, pan2014ultra}. The modulation of electronic, magnetic, and mechanical properties through curvature provides a promising avenue for the development of nanoresonators, sensors, and spintronic devices.

In this work, we present a first-principles study on the structural and electronic properties of lateral biphenylene/graphene heterojunctions (\bpng). We focus on: (i) the stability and electronic structure of \bpng\, as a function of the graphene region width, G$_\text{n}$; (ii) the effects of hydrogenation on \bpngn, leading to the formation of catenary-like rippled graphene stripes [\bpngh]; and (iii) the electronic transport properties of \bpngh\, under mechanical strain.

\begin{figure}
        \includegraphics[width=7.5cm]{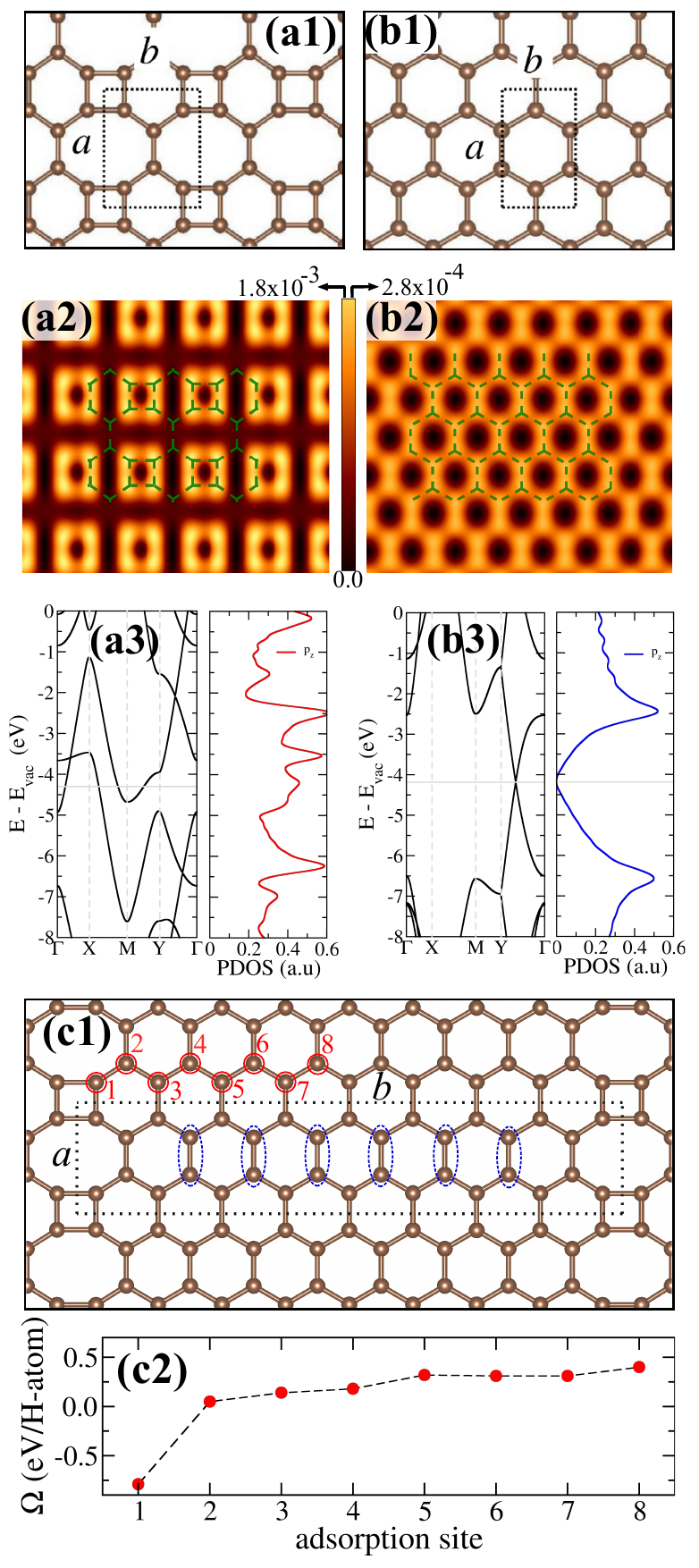}
        \caption{\label{fig:model0} Structural models, simulated STM images (empty states, $E_F+1$\,eV), electronic band structure, and the projected density of states of biphenylene (a1)-(a3) and graphene (b1)-(b3). (c1) Structural model o \bpng$_6$, and (c2) the formation energy of (\bpng$_6$)$_\text{1H}$  ($\Omega_\text{1H}$) as a function of the hydrogen adsorption site, 1\,-\,8 as indicate in (c1). The STM images are made with isovalue of [$e^{-}/\AA^{2}$.]}
        %The zero energy was set to the vacuum level (E$_\text{vac}$).}  
\end{figure}

\section{COMPUTATIONAL DETAILS}

\begin{figure*}
        \includegraphics[width=15cm]{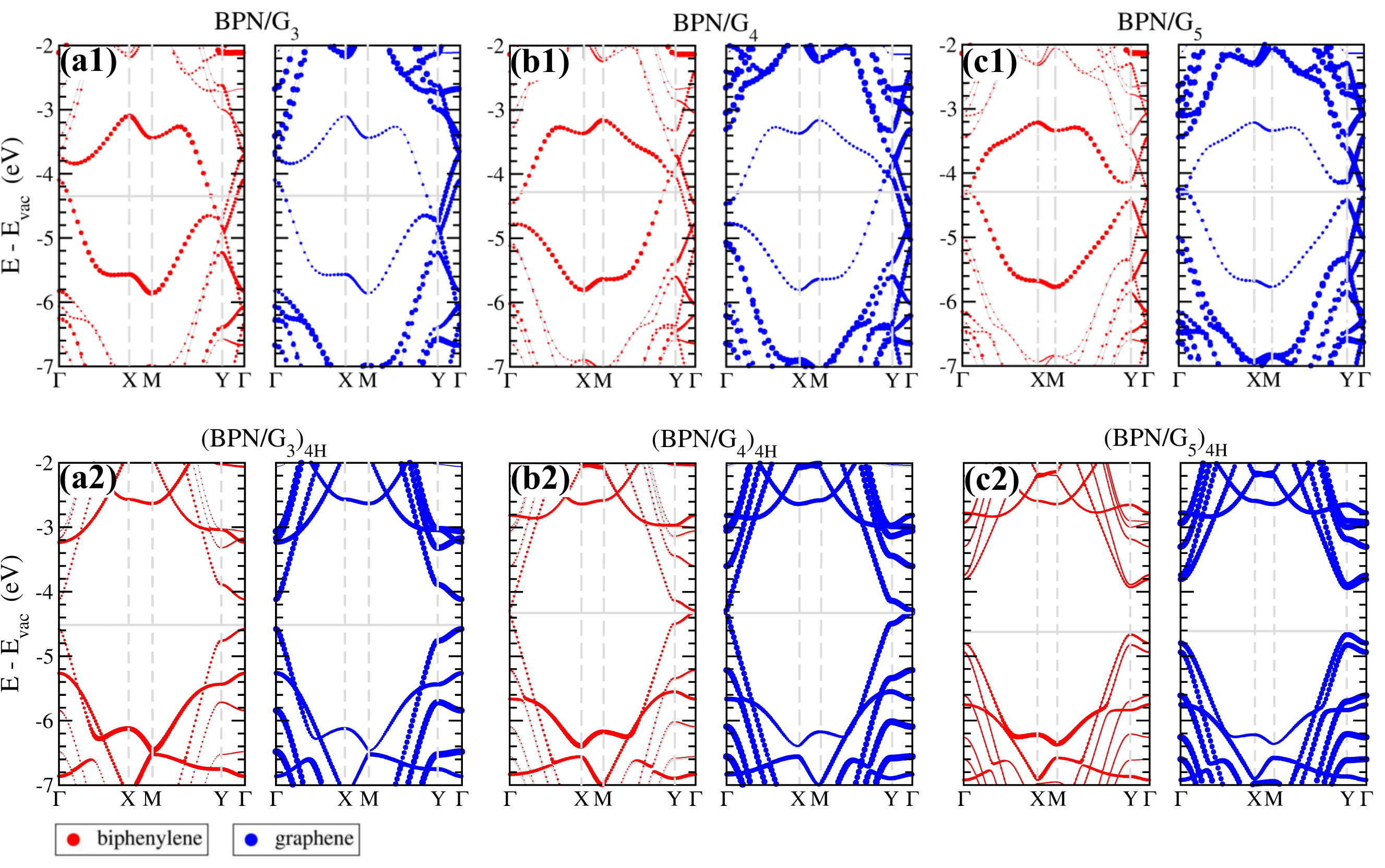}
        \caption{\label{fig:band} Orbital projected electronic band structures of \bpngn\, and \bpngh\, with n\,=\,3\,(a1)-(a2), 4\,(b1)-(b2), and 5\,(c1)-(c2). The density of the orbital projection is proportional to the size of the filled circles. Red (blue) circles indicate the projection on the BPN (G$_\text{n}$) region.}
\end{figure*}

In this study, we employed Density Functional Theory (DFT) calculations to investigate the properties of biphenylene-graphene lateral heterosctructures.
All calculations were performed using the computational software VASP (Vienna Ab initio Simulation Package) \cite{Kresse1993}. We utilized Projector-Augmented Wave (PAW) \cite{Blchl1994} potentials for an accurate description of electronic interactions. The selected exchange-correlation functional was GGA-PBE (Generalized Gradient Approximation - Perdew, Burke, and Ernzerhof)\cite{Perdew1996}.
For the plane-wave basis set, we set a kinetic energy cutoff of 450 eV, ensuring the proper inclusion of high-energy contributions in our calculations. Brillouin zone sampling was carried out using a k-point grid centered at the $\Gamma$-point, following the Monkhorst-Pack scheme \cite{Monkhorst1976}, with an 8$\times$2$\times$1 mesh. The $k$-mesh choice was supported by convergence analysis. Ionic relaxation was performed using the conjugate gradient algorithm, with all atoms in the supercell relaxed until atomic forces were smaller than $0.01$ eV/$\AA$.

The transport calculation was carried out using the Spanish Initiative for Electronic Simulations with Thousands of Atoms (\textsc{Siesta}) code\,\cite{Soler2002,Papior2017}, which combines DFT theory with non-equilibrium Green’s function (NEGF) methods. The core electrons were described by norm-conserving Troullier-Martins pseudopotentials\,\cite{Hamann1979} which are available through the \textsc{Pseudo-Dojo} project\,\cite{Garca2018}, while the valence electrons were treated using a Double-Zeta Polarized (DZP) basis set of numerical atomic orbitals (NAOs)\cite{Junquera2001}, with an energy shift of 0.02 Ry and a real-space mesh cutoff of 300 Ry. A reciprocal space Monkhorst-Pack scheme with a 240$\times$1$\times$100 (10$\times$1$\times$100) for the electrode and 240$\times$1$\times$1 (10$\times$1$\times$1) for the scattering were used for the wave (tile) directions. To calculate the transmission function [$T(E)$] and current density ($I$) a 480$\times$1$\times$1 (20$\times$1$\times$1) k-points mesh was used for the wave (tile) direction.

For the non-stressed systems, the total transmission coefficients, \( T(E,V) \) were calculated self-consistently at finite bias and integrated according to the Landauer–Büttiker scheme to
provide the current density given by
\begin{equation}
   \small I(V) = \dfrac{2 e}{A h} \int_{-\infty}^{+\infty} T(E,V) \left[ f(E,\mu_\text{L}) - f(E,\mu_\text{R}) \right] dE,
\end{equation}
where $e$ is the electron charge, $h$ is Planck's constant, $A$ the surface area perpendicular to the transport direction, and $f(E,\mu)$ the Fermi–Dirac function. An applied bias, $V$, shift the left and right chemical potential as $\mu_{L/R}=E_{F}\pm eV/2$, with $E_{F}$ being the electrode Fermi energy, and the electronic temperature $T = 300$\,K.

\section{Results and discussions}

\subsection{Pristine biphenylene and graphene}

Let us start examining the structural and electronic properties of the pristine systems, namely biphenylene (BPN) and graphene (G), to check the consistency and reliability of our results compared with previous studies\,\cite{Fan2021, luo2021first, kuritza2024directional}. 

In Fig.\,\ref{fig:model0}(a1) and (b1) we present the structural models of biphenylene (BPN) and graphene (G). At the equilibrium geometry, the lattice parameters of the rectangular unit cell of BPN are $a$\,=\,4.52 and $b$\,=\,3.76\,\AA, and using a rectangular unit cell to describe the periodicity of graphene, we found $a$\,=\,4.26 and $b$\,=\,2.47\,\AA. Regarding the energetic properties, our results of cohesion energy ($E^c$) of BPN  and graphene, $E^c$ of 7.43 and 9.19\,eV/atom, are in good agreement with the current literature\,\cite{Luo2021,Ma2022}. The carbon atoms in BPN are three-fold coordinated, as in graphene; however, instead of only hexagonal rings, in BPN the carbon atoms form a combination of square, hexagonal, and octagonal rings, which results in a distinct distribution of the surface electronic states. This is what we found in the simulated STM images [Fig.\,\ref{fig:model0}(a2) and (b2)], namely, (i) the lattice structure of BPN is characterized by the formation of bright spots lying on the carbon atoms that form the square rings, while (ii) in graphene the bright spots form a uniform hexagonal structure.   In the former [(i)], the electronic density of states close to the Fermi level increases as a result of the distortion of the C$_3$ symmetry of the sp$^2$ bonds, which in turn leads to a more reactive site on the BPN surface\,\cite{lahiri2010extended, brito2011hydrogenated}. 

It is noteworthy that BPN presents a type-II Dirac cone above the Fermi level ($E_F$), along the $\Gamma$-X direction [Fig.\,\ref{fig:model0}(a3)]; while due to the folding of the Brillouin zone (BZ), the linear energy bands of graphene form the Dirac point (DP) along the Y-$\Gamma$, Fig.\,\ref{fig:model0}(b3). Further orbital projected density of states (PDOS) reveals that the energy bands near the Fermi level mostly comprise the out-of-plane C-p$_z$ orbitals, and both systems present somewhat similar values of work function ($\Phi$), i.e., 4.30 and 4.20\,eV for BPN and G, respectively\,\cite{yu2009tuning, garg2014work, leenaerts2016work, Bafekry_2021}.  
 
\subsection{\bpng\, heterostructures}

\begin{figure}
	\subfigure{\includegraphics[width=\columnwidth]{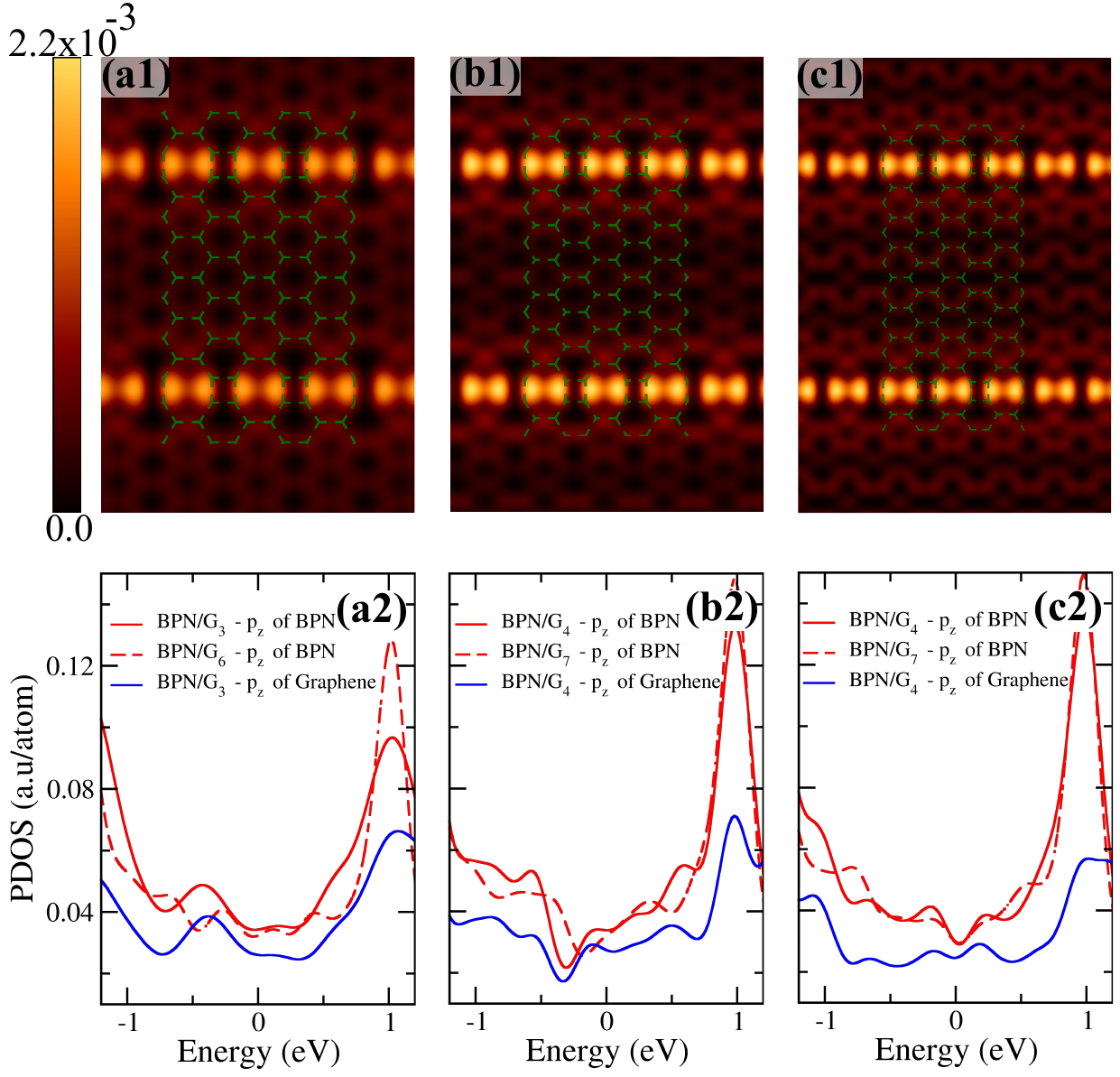}}
	\caption{Simulated STM images of the  unoccupied states ($E_F+1$\,eV) of \bpngn\, with n\,=\,3\,(a1), 4\,(b1), and 5\,(c1), and the projected density of states of \bpngn\, with n\,=\,3 and 6\,(a2), 4 and 7\,(b2), and 5 and 8\,(c2). The orbital projections on the BPN and G$_\text{n}$ regions are indicated by red and blue lines. denotes with isovalue of [$e^{-}/\AA^{2}$.] }
	\label{fig:stm}
\end{figure}

In this study, we propose a BNP/G heterostructure, which can be thought of as biphenylene and graphene stitched along the $a$ direction of Fig.\ref{fig:model0}(a1) and (b1). Our focus has been on heterostructures that consist of single BPN stripes between periodic arrays of graphene (armchair) ribbon structure, \bpngn. The width of the ribbons is based on the number (n) of C-dimer lines that separate the BPNs. To illustrate this, in Fig.\,\ref{fig:model0}(c1) we present a \bpngn\, heterostructure composed of six C-dimer lines in between the BPN stripes, \bpng$_6$.

The energetic stability of BPN/G$_\text{n}$, with n\,=\,3-8, was examined through the calculation of the cohesion energy, $E^c$, which is defined as the total energy comparison between the final system, BPN/G$_\text{n}$ ($E[\text{BPN/G}_\text{n}]$), with the ones of the isolated components, C atoms ($E[\text{C}]$),
$$
E^c = \text{N}_\text{C}\times E[\text{C}] - E[\text{BPN/G}_\text{n}],
$$
N$_\text{C}$ is the number of C atoms in BPN/G$_\text{n}$. Our results of $E^c$, summarized in Table\,\ref{tab:energy}, reveal that (as expected) the cohesion energy of BPN/G$_\text{n}$ tends to the one of graphene (9.19\,eV/atom) as n increases, i.e., $E^\text{c}$\,=\,9.06\,$\rightarrow$\,9.13\,eV for n\,=\,3\,$\rightarrow$\,8\,\cite{hong2024design}. At the equilibrium geometry, BPN and G remain flat. Due to the lattice mismatch in the $a$-direction, the graphene C-C bond lengths at the BPN-G interface are stretched by $\sim$\,0.02\,\AA\, in comparison with that of pristine graphene. Changes in the bond angles, on the other hand, preserve the planar sp$^2$ hybridizations by allowing strain relief at the interface. Further phonon spectra calculations confirm the dynamical stability of \bpng$_6$, Fig.\,SMS1.

{\renewcommand{\arraystretch}{1.2}
\begin{table}
\caption{\label{tab:energy} Equilibrium lattice constants ($a$ and $b$), cohesion energy ($E^c$ in eV/C-atom), and work function ($\Phi$ in eV) of the pristine systems, BPN and G, and \bpngn\, heterostructure.}
\begin{ruledtabular}
    \begin{tabular}{crrrc}
       Structure & $a$\,(\AA) & $b$\,(\AA)&   $E^c$ & $\Phi$\\
       \hline
       Biphenylene & 4.52 &  3.76   & 7.43 & 4.30\\
      Graphene &    4.26 & 2.47  & 9.19  & 4.20 \\
      \hline
       BPN/G$_3$    & 4.32 &  13.68
    & 9.06 & 4.34 \\
      BPN/G$_4$ &  4.31 &16.14
 & 9.07  & 4.28\\
       BPN/G$_5$ & 4.31 & 18.64 &  9.10 &  4.29\\
       BPN/G$_6$ & 4.30 &21.09
  & 9.11  & 4.33 \\
     BPN/G$_7$ & 4.30 & 23.57
   & 9.12  & 4.28 \\
     BPN/G$_8$ & 4.30 & 26.04
  & 9.13   & 4.29 \\
    \end{tabular}
\end{ruledtabular}
\end{table}
}

The electronic band structures for \bpngn\, systems with n=3,\,4, and 5 are presented in Fig.\,\ref{fig:band}(a1)-(c1). The orbital projections reveal that, near the Fermi level ($E_F$), the electronic states with wave-vector parallel to the BPN-G$_\text{n}$ interface, $\Gamma$-X direction,  are mostly concentrated on the BPN stripes (red circles in Fig.\,\ref{fig:band}), while the electronic states with wave-vector along the $\Gamma$-Y direction (i.e., perpendicular to the interface)  are mostly concentrated in the G$_\text{n}$ region (blue circles in Fig.\,\ref{fig:band}). Notably, the dispersion features of the energy bands reflect the folding effect in the Brillouin zone due to the periodic perturbative potential produced by the BPN stripes ($V_\text{BPN}$). Here, the periodicity of $V_\text{BPN}$ is given by the number of C-dimers (n) in G$_\text{n}$, which in its turn can be classified into three groups: n = 3p, 3p+1, and 3p+2 (with p\,=\,integer). In this case, \bpngn\, structures with n\,=\,3, 4, and 5 correspond to p\,=\,1. In Fig\,SMS2 appendix, we present the electronic band structures of \bpngn\, with n\,=\,6, 7, and 8, corresponding to p\,=\,2, showing that \bpngn\, structures belonging to the same group present similar electronic band structures. 

In general, the presence of linear defects in graphene, like grain boundaries, leads to an increase of the local density of states near the Fermi level, which can be identified through STM images by the formation of bright lines along the defective regions\,\cite{simonis2002stm, lahiri2010extended, koepke2013atomic}. Here, by using the Tersoff-Hamann approximation\,\cite{tersoff1985theory}, we simulate the STM images of \bpngn. In Fig.\,\ref{fig:stm}(a1)-(c1) we present the simulated images of the unoccupied states, $E_F+1$\,eV, where we can identify the formation of bright stripes lying on the C atoms that form the square rings of BPN. Since at the equilibrium geometries the \bpngn\, heterostructures are planar, we can infer that the surface local density of states will dictate the features of the STM images, which can be examined in light of the projected density of states (PDOS) near the Fermi level, Fig.\,\ref{fig:stm}(a2)-(c2). As expected, the density of states projected onto the BPN region (red lines) is larger compared with that on the G$_\text{n}$ region (blue lines). Similarly, when we consider the occupied states, $E_\text{F}-1$\,eV, we can infer that the STM images will also be characterized by bright spots along the BPN stripes.

\begin{figure}
	\subfigure{\includegraphics[width=\columnwidth]{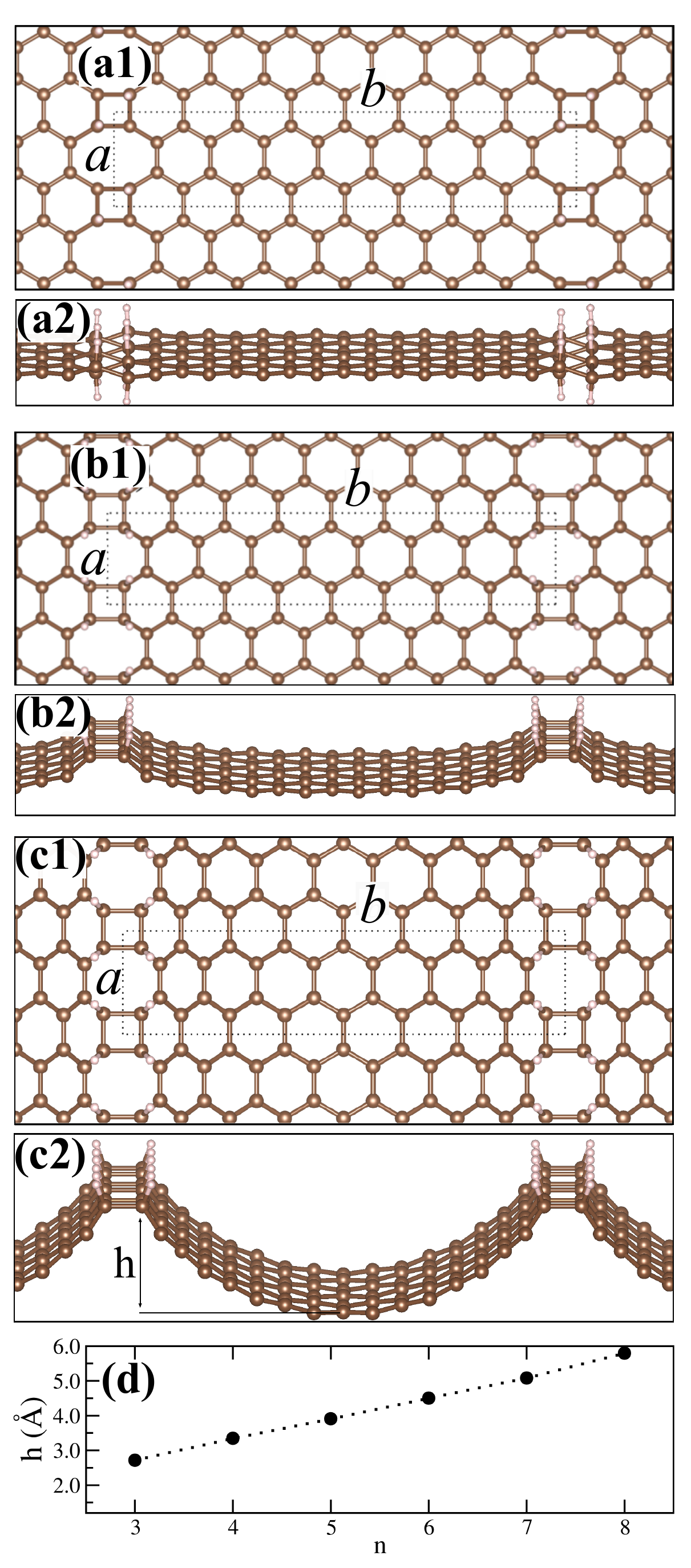}}
	\caption{Structural models of hydrogenated (BPN/G$_6$)$_\text{4H}$ with H adatom on  the opposite sites of the BPN's square ring (a1)-(a2), the same side of the square ring at the planar (b1)-(b2), and curved   (c1)-(c2) G$_6$ region. (d) Curvature height h as a function of the number of G-units n.} 
	\label{fig:model1}
\end{figure}

\subsection{Hydrogenated \bpngn}

\subsubsection{Structural and electronic properties}

Nanopattering is one very promising path to perform material engineering in 2D platforms, in particular, the ones based on hydrogen adatoms\,\cite{boukhvalov2008hydrogen, whitener2018hydrogenated, cadelano2010elastic, gao2011band}. For instance, the emergence of magnetic properties in graphene/graphane lateral heterostructures\,\cite{singh2009electronics} and the formation of one-dimensional semiconducting channels guided by lines of hydrogen adatoms in graphene\,\cite{lian2016electronic}. The hydrogenation of BPN has been the subject of extensive research in recent years\,\cite{liao2021new, lee2021band, xie2022effective, chen2023structural}, addressing both the physical and chemical understanding of these novel materials as well as (potential) technological applications\,\cite{zhang2021intrinsic, demirci2022hydrogenated, zhang2024buckling}. 

We proceed with our investigation by incorporating hydrogen atoms into the \bpngn\, heterostructure, \bpngn\,$\rightarrow$\,(\bpngn)$_\text{H}$. The energetic stability of the H adatoms was examined in light of the calculation of the formation energy ($\Omega_{N\text{H}}$) using the following equation,
\begin{equation}
    \Omega_{N\text{H}}=E[(\text{BPN/G}_\text{n})_{N\text{H}}] -\left(E[\text{BPN/G}_\text{n}] + \frac{N}{2}\mu_{\rm H_2}\right),
\end{equation}
where $E[\text{(BPN/G}_\text{n})_{N\text{H}}]$ represents the total energy of \bpngn\ adsorbed with $N$ H atoms per unit-cell, $E[\text{BPN/G}_\text{n}]$  denotes its pristine counterpart, and $\mu_{\rm H_2}$ stands for the total energy of an isolated H$_2$ molecule. We use the BPN/G$_6$ system to introduce the H adatom into various sites, as illustrated in Fig.\,\ref{fig:model0}(c1). Our findings reveal that the exothermic incorporation of H atoms takes place only at the BPN region above the carbon atoms in the square ring of BPN, $\Omega_\text{1H}$\,=\,$-0.79$\,eV/H-atom [Fig.\,\ref{fig:model0}(c2)], whereas,  as the position of hydrogen moves away from biphenylene, going deeper into the G$_6$ region (site 1\,$\rightarrow$\,8) the H adsorption becomes endothermic, $\Omega_\text{1H} > 0$, with $\Omega_\text{1H}$\,=\,$0.40$\,eV/H-atom far from the BPN stripe. Indeed, the energetic preference for the hydrogen incorporation along the extended defects in graphene  has been already theoretically predicted based on first-principles DFT calculations\,\cite{brito2011hydrogenated}.

{\renewcommand{\arraystretch}{1.2}
\begin{table}[]
\caption{\label{tab:energy} Formation energy of (\bpngn)$_\text{4H}$ and the relative decrease in the  lattice vector perpendicular to the BPN stripes ($\Delta b/b_0$).}
\begin{ruledtabular}
    \begin{tabular}{lcc}
       Structure & $\Omega_\text{4H}$ (eV/H-atom)  &   $\Delta b/b_0$ (\%)\\
       \hline
       BPN/G$_3$  &     $-1.30$        &      10.2   \\
       BPN/G$_4$  &     $-1.35$        &      11.4   \\
       BPN/G$_5$  &     $-1.37$        &      12.0   \\
       BPN/G$_6$  &     $-1.39$        &      12.7   \\
       BPN/G$_7$  &     $-1.41$        &      13.1   \\
       BPN/G$_8$  &     $-1.43$        &      14.1   \\
    \end{tabular}
\end{ruledtabular}
\end{table}
}

After establishing an energetic preference for H atoms adsorbed within the BPN regions, we proceed to specifically place four H adatoms onto the C atoms forming the square rings of the BPN/G$_6$ structures, denoted as (BPN/G$_6$)$_{4\text{H}}$. We have considered two structural models, one consisting of H atoms sitting on opposite sites and another with H atoms lying on the same (nearest neighbor) C atoms of the square ring, Fig.\,\ref{fig:model1}(a) and (b). The former configuration is somewhat close to that proposed by Demirci {\it et al.}\,\cite{demirci2022hydrogenated} for half-hydrogenated BNP. Here, we have BPN's stripes embedded in graphene (G$_\text{n}$) rather than pristine BPN, which results in a different picture of the energy stability of H adatoms. In this case, we found that the energetic preference for H atoms positioned on the same side of the square ring of C atoms [Fig.\,\ref{fig:model1}(b)] leads to a barrierless bending process of the G$_6$ region [Fig.\,\ref{fig:model1}(b)\,$\rightarrow$\,(c)] dictated by the sp$^3$ hybridization of the hydrogenated C atoms. At the equilibrium geometry, the lattice vector $b$ contracts by about 13\,\% ($\Delta b$\,=\,13\,\%), giving rise to a catenary-like rippled graphene structure, side view in Fig.\,\ref{fig:model1}(c2).

The results of our formation energy for the other (\bpngn)$_\text{4H}$ structures, and the relative decrease in the vector perpendicular to the BPN stripes ($\Delta b/b_0$), as a result of the curvature effect, are shown in Table\,II. In Fig.\,\ref{fig:model1}(c1)-(c2) and (d), we present the equilibrium geometry of the rippled \bpngh\, structure and magnitude of the curvature (h) in each case, which is determined by considering the difference between the highest and lowest carbon heights. We observe a nearly linear increase in curvature as a function of the G-units, ranging from h\,=\,2.9\,\AA\, for n\,=\,3 to h\,=\,5.8\,\AA\, for n\,=\,8.

Focusing on the electronic structure of \bpngh, we find that the metallic bands present in \bpngn\, are washed out upon hydrogenation, Fig.\,\ref{fig:band}(a1)-(c1)\,$\rightarrow$\,(a2)-(c2), and  the \bpngh\, heterostructures become semiconductors. The orbital projection of the energy bands near the Fermi level reveals the predominant contributions from the graphene regions, G$_\text{n}$, where the formation of hydrogenated BPN stripes leads to quantum-confinement effects of the electronic states. In this context, \bpngh\, heterostructures can be viewed as a set of rippled armchair graphene nanoribbons (GNRs).  In fact, when we examine the band gaps ($E_\text{g}$) of \bpngh\, as a function of the G$_\text{n}$ width, see Table\,III, we find that values $E_\text{g}$ follow the 3p, 3p+1, and 3p+2 classification, for integer values of p, dictated by the (particular) features of orbital hybridizations of the edge states as predicted by Son et al.\cite{son2006energy,note1}. 
In Fig.\,SMS2 we present the electronic band structures of \bpngh, with n\,=\,6, 7, and 8, which corresponds to p\,=\,2.

{\renewcommand{\arraystretch}{1.2}
\begin{table}[]
\begin{ruledtabular}
\caption{\label{tab:BG} Bandgap  (\bpngh)}
    \begin{tabular}{ccc}
       \multicolumn{3}{c}{\bpngh} \\
       \hline
       G$_\text{n} $ & $\Gamma$-point (eV) & Y-point (eV) \\
       \hline
       3  &  0.46 & - \\ 
       4  &  0.03 & - \\
       5  &  -    & 0.73 \\
       6  & 0.34  & - \\
       7  & 0.02  & - \\
       8  & -     & 0.51 \\
%       9  & 0.27  & - \\
%       10 & 0.01  & - \\
    \end{tabular}
\end{ruledtabular}
\end{table}
}

\subsubsection{Compressive and tensile strain}

Curved structures in graphene\,\cite{ortolani2012folded} and graphene nanoribbons\,\cite{van2015bending} offer a new and important degree of freedom for the design of electromechanical nanodevices\,\cite{qi2011enhanced,   wang2019advanced, miao2023flexible}, like strain sensors made by graphene ripples\,\cite{wang2011super}; and proposals for application in catalytic processes\,\cite{qu2018effect,liu2020substrate}.  In this context, understanding the effect of curvature on the electronic properties of graphene is an important issue to be addressed\,\cite{zhang2010influence, ortolani2012folded, hammouri2017ab}. In the previous section, we showed that selective hydrogenation of \bpngn\ results in a periodic array of catenary-like ripplings in \bpngh, followed by quantum-confinement effects, and metal\,$\rightarrow$\,semiconductor transition. In the sequence, we will examine the effect of the mechanical strain ($\Delta$) in the structural, and electronic properties of \bpngh.

The energetic cost to stretch/compress ($\Delta$\,$>$\,0/$\Delta$\,$<$\,0) the \bpngh\, heterostructure can be inferred through the calculation of the bending energy ($E^\text{bend}$) defined as the following total energy difference, 

\begin{equation}
E^\text{bend} = \frac{1}{N_\text{C}}\left(E^s[(\text{BNP/G}_\text{n})_\text{4H}] - E^0[(\text{BNP/G}_\text{n})_\text{4H}]\right),
\end{equation}
where $E^s[(\text{BNP/G}_\text{n})_\text{4H}]$ and $E^0[(\text{BNP/G}_\text{n})_\text{4H}]$ are the total energies of the strained and ground state \bpngh\, for n\,=\,3-8, and $N_\textbf{C}$ is the number of C atoms in G$_\text{n}$, i.e., $N_\text{C}$\,=\,2$\times$n. 

\begin{figure}[H]
        \includegraphics[width=\columnwidth]{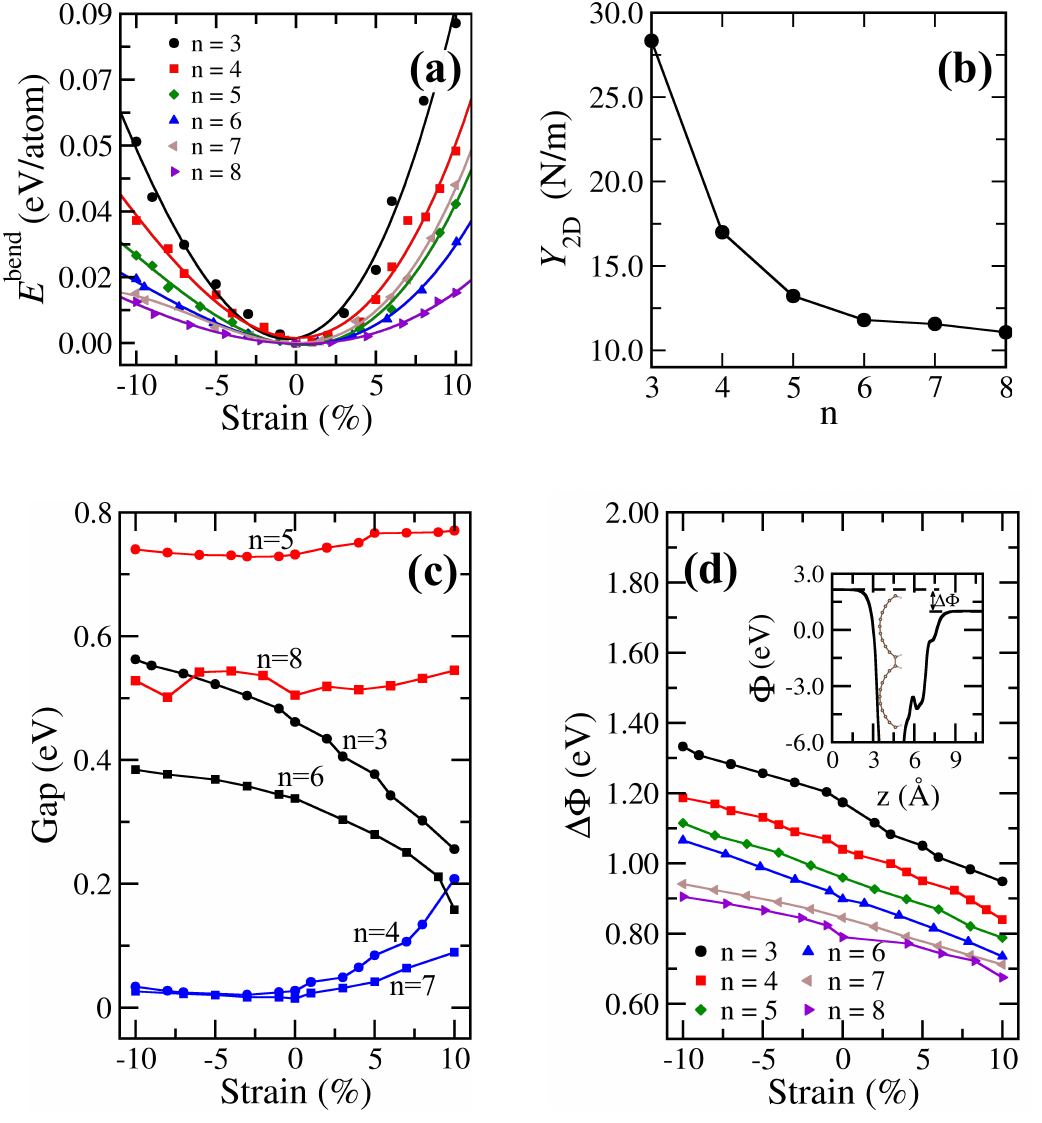}
        \caption{\label{fig:strain}(a) Bending energy ($E^\text{bend}$), (b) Young's modulus of \bpngh, (c) the gap energy, and (d) internal dipole ($\Delta\Phi$, schematically shown in the inset) as a function of strain. In (a), the calculated values, using eq.\,(3), are indicated by the filled symbols, and the solid line indicates the third-order polynomial fitting.}
\end{figure}

Our results of $E^\text{bend}$ for $\Delta$ up to 10\,\%, summarized in Fig.\,\ref{fig:strain}(a), reveal that the deformation energies in the \bpngh\, structures are comparable with those observed in the radial deformation of carbon nanotubes \cite{akinwande2017review}.  Moreover, it should be noted that the asymmetric feature of the bending energy as a function of the strain can be, accordingly, captured by a third-order polynomial fit. As expected, the stiffness of \bpngh\, is much lower compared with that of graphene, which can be verified through the calculation of the 2D Young's modulus ($Y^\text{2D}$)\,\cite{akinwande2017review}. We found $Y^\text{2D}$ of 345\,N/m for single-layer graphene, which is in good agreement with the experimental measurement of  340\,N/m\,\cite{lee2008measurement}; whereas $Y^\text{2D}$ reduces by more than 10-fold in the \bpngh\, systems, reaching lower values for higher values of n, namely $Y^\text{2D}$\,$\approx$\,27\,$\rightarrow$\,12\,N/m for n\,=\,3\,$\rightarrow$\,8, Fig.\,\ref{fig:strain}(b).

According to the authors in Ref.\,\cite{son2006energy}, the atomic relaxations at the ribbon edge sites, specifically the C-C bond lengths, determine the band gap dependency with the width of the GNRs. This, in turn, leads to changes in the hopping integrals between the C-p$_z$ orbitals. Indeed, as shown in Fig.\,\ref{fig:strain}(c), this is what we found in the strained \bpngh\, heterostructures. For example, upon stretching ($\Delta>0$), the increase of the C-C bond lengths at the edge sites in G$_\text{n}$ promotes the decrease (increase) of the band gaps of \bpngh\, with n=3 and 6 (n=4 and 7); meanwhile we find that the band gaps of (\bpng$_{5,\,8}$)$_\text{4H}$ are slightly affected upon strain.

Although this is not the main focus of our study, it is worth pointing out that, different from the waved graphene structures, we find that the catenary-like rippled geometry leads to the emergence of an out-of-plane dipole that can be mechanically tuned. As depicted in Fig.\,\ref{fig:strain}(d), the electric polarization is normal to the \bpngh\, sheet, suggesting that the curved topology of \bpngh\, can be employed for applications in stretchable energy harvesting devices and (nano)sensors\,\cite{qi2011enhanced, ahmadpoor2015flexoelectricity, huang2018assembly}. We believe that further investigations are worth doing in order to explore such a mechanically tunable polarization in curved 2D systems.

% Electronic transport 

%\begin{itemize}
%{Electronic transport: some references} 
%    \item Conductance of Curved GNRs J. Phys. Chem. C 123, 21805 (2019)\cite{zhang2019conductance}
%    \item Current-voltage (I-V) characteristics, PRB 81, 205437 (2010)\,\cite{topsakal2010current}
 %   \item Voltage-dependent conductance of a single GNR, Nature Nanotech. 7, 713 (2012)\,\cite{koch2012voltage}
 %   \item Ab initio study of the electronic, Physica E 89, 170 (2017)\,\cite{hammouri2017ab}
%\end{itemize}

\subsubsection{Electronic Transport and Current-Voltage Characteristics}

\begin{figure}
    \includegraphics[width=\columnwidth]{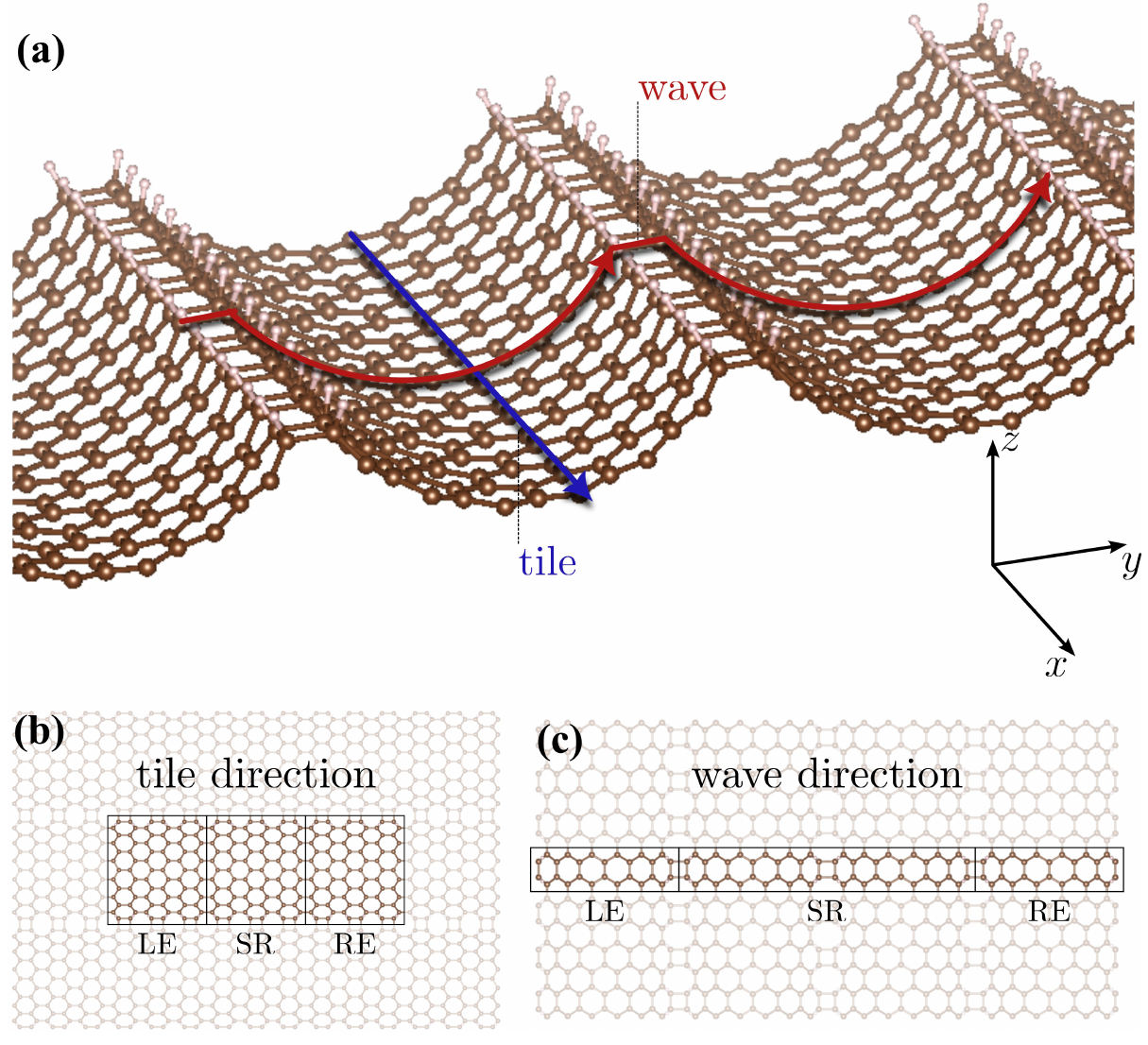}
    \caption{\label{fig:tran-1} Electronic transport configuration for the BPN-G$_n$ heterostructure along the tile (blue) and wave (red) directions. The inset in the top-left corner illustrates the electrode (LE and RE) and the scattering region (SR) used for both directions.}
\end{figure}

In order to provide a more complete view of the applicability of the catenary-like rippled graphene stripes in the design of nanodevices, the electronic transport properties of the \bpngh\, heterostructures were investigated by calculating the electronic transmission function [$T(E)$] for systems with n\,=\,3, 4, and 5, in both directions, named tile and wave, as illustrated in Fig.\,\ref{fig:tran-1}(a). The setup includes a scattering region (SR) connected to two semi-infinite electrodes, referred to as the left electrode (LE) and right electrode (RE). In the tile configuration [Fig.\,\ref{fig:tran-1}(b)], 9 unit cells of \bpngh\, were used (6 for the electrodes and 3 for the SR), while in the wave configuration [Fig.\,\ref{fig:tran-1}(c)], the setup consisted of 4 \bpngh\, unit cells (2 cells for the electrodes and 2 cells for the SR). For the calculation of $T(E)$, we have considered an energy range of $\pm$\,1\,eV with respect to the Fermi level; thus, as shown in the orbital-projected band structure [Fig.\,\ref{fig:band}(a2)-(c2)], the transmission channels are mostly ruled by the C-$2p_z$ orbitals of the G$_\text{n}$ stripes.

\begin{figure}
    \includegraphics[width=\columnwidth]{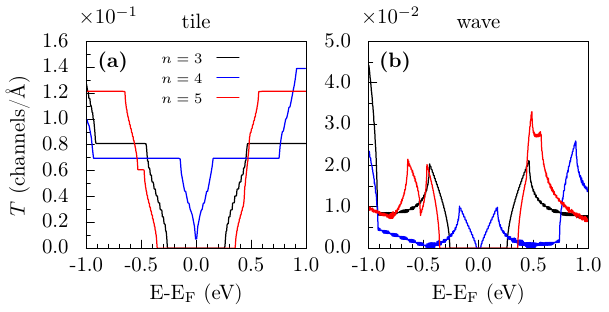}
    \caption{\label{fig:tran-2} Electronic transmission [$T(E)$] for $n=3,4,5$ configurations along the (a) wave and (b) tile directions.} 
\end{figure}

Our results of $T(E)$, depicted in Fig.\,\ref{fig:tran-2}, are (i) consistent with the semiconducting character of \bpngh, with $E_g$\,=\,0.46 and 0.73\,eV for n=3 and 5, and the nearly semi-metallic character ($E_g$\,=\,0.03\,eV) for n\,=\,4; and (ii) confirm the emergence of the anisotropic electronic transport as inferred from electronic band structures [Fig.\,\ref{fig:band}]. We found that the transmission function along the tile direction ($T_\text{tile}$) is about an order of magnitude larger than that along the wave direction ($T_\text{wave}$), as shown in Fig.\,\ref{fig:tran-2}(a) and (b). It is noteworthy that the features of $T_\text{tile}$, characterized by the presence of transmission plateaus, somewhat mimic the ones of armchair GNRs. In contrast, in $T_\text{wave}$ we find the emergence of peaks and valleys due to the hydrogenated BPN regions acting as scattering centers.

\begin{figure}
    \includegraphics[width=\columnwidth]{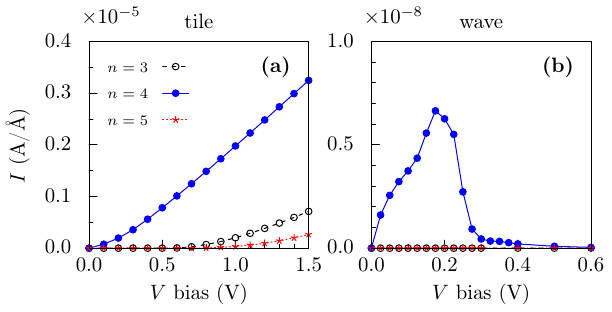}
    \caption{\label{fig:current-voltage} Current density ($I$) as a function of bias voltage for the tile (a) and wave (b) directions.}
\end{figure}

Using the Landauer-Büttiker scheme [Eq. (1)], we calculate the electronic current density  along the tile ($I_\text{tile}$) and wave directions ($I_\text{wave}$) as a function of the applied bias voltage ($V$). We observe a quasi-linear behavior for $I_\text{tile}$, which initially exhibits a parabolic increase with $V$ around 0.6, 0.3, and 0.7 $V$ for $n=$3, 4, and 5, followed by a linear growth [Fig. \ref{fig:current-voltage}(a)]. The behavior observed in this case results from the 'V' shaped configuration presented in the transmittance $T_\text{tile}$, while the linear behavior of $I_\text{tile}$ can be attributed to the plateaus with constant transmittance, Fig.\ref{fig:tran-2}(a). In contrast, along the wave direction [Fig. \ref{fig:current-voltage}(b)] for $n$\,=\,3, and 5 the current density $I_\text{wave}$ is zero, as expected for a gaped material (semiconductor), while for $n$\,=\,4, we observe the occurrence of negative differential resistance (NDR), characterized by a decrease in current as the applied bias voltage increases beyond 0.17\,V. 

The NDR effect,  is characterized by an initial increase followed by a reduction of the electronic current as a function of the applied bias voltage. This behavior is attributed to the alignment and subsequent misalignment of electronic states between the electrodes as the applied bias increases. Here, the emergence of the NDR for $I_\text{tile}$ can be understood by looking at the electronic band structures of \bpngh\, Fig.\,\ref{fig:band}(a2)-(c2), especially the energy bands along the Y-$\Gamma$ direction, that is, the ones with wave vectors parallel to the tile direction. Let us look at occupied (unoccupied) energy bands of (\bpng$_4$)$_\text{4H}$, as shown in Fig.\,\ref{fig:band}(b2). In this instance, when the applied bias voltage, $V$, is smaller than the bandwidth of the electrodes occupied (LE) and unoccupied (RE) states, the energy levels are resonant. As a result, the band-to-band (LE\,$\rightarrow$\,RE) tunneling process can occur through the scattering region, leading to the increase of the electronic current as a function of the bias voltage. The increase of the bias potential may lead to the reduction of the band alignment between the occupied (LE) and empty (RE) electronic states. For example, for $V>0.17$\,V in (\bpng$_4$)$_\text{4H}$, and further increase of $V$ can result in a whole band misalignment and, consequently, the suppression of band-to-band tunneling process, LE\,$\not\rightarrow$\,RE. Further discussion on the NDR, detailing the contribution of the transmittance coeficient, $T(V)$, and the Fermi-Dirac distribution of the electrodes  has been presented in the Supplemental Material. 

A quantitative measure of the NDR's strength can be extracted from the respective peak-to-valley ratio (PVR). As summarized in  Table\,\ref{tab:pv_ratio}, we find that (\bpng$_4$)$_\text{4H}$ presents the strongest NDR, with a PVR of 179.3, highlighting its potential for high-frequency oscillators and memory devices\cite{Wu2012,Tiwari2019,Bhattacharya2021}.

{\renewcommand{\arraystretch}{1.2}
\begin{table}[h!]
\caption{\label{tab:pv_ratio} Peak-to-valley ratio (PVR) for the wave configuration in the BPN/G$_n$ heterostructures.}
\begin{ruledtabular}
\begin{tabular}{cccc}
 (\bpngn)$_\text{4H}$ & PVR & $V_\text{p}$ (V) & $V_\text{v}$ (V) \\
\hline
 n\,=\,3 & 3.2 & 0.10 & 0.20 \\
 n\,=\,4 & 179.3 & 0.17 & 0.60 \\
 n\,=\,5 & 8.1 & 0.05 & 0.15  \\
\end{tabular}
\end{ruledtabular}
\end{table}
}

\begin{figure}[h!]
    \includegraphics[scale=1]{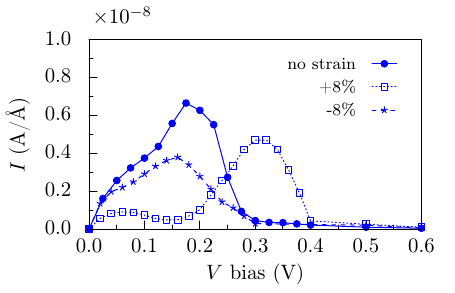}
    \caption{\label{fig:current-strain} Electronic current ($I$) in the wave direction for the (BPN/G$_4$)$_\text{4H}$ system under tensile (+8\%) and compressive (-8\%) strain.}
\end{figure}

The investigation of negative differential resistance (NDR) in these systems underscores the role of external factors, such as strain, in modulating electronic transport properties. For the unstrained system, the observed NDR behavior arises from the alignment between transmission peaks and the Fermi-level distribution, as illustrated in Fig\,[SMS4]. Introducing strain adds an extra dimension of tunability. Compressive strain, depicted in Fig.,\ref{fig:current-strain}, primarily decreases the peak-to-valley ratio (PVR) by modifying the electronic coupling and overlap, while largely preserving the band structure. Conversely, tensile strain introduces more substantial effects, including the emergence of distinct NDR regions, as detailed in Table \ref{tab:pv_ratio-str}.

Figure \ref{fig:current-strain} demonstrates how compressive and tensile strain affect the electronic current $I_\text{wave}$ of the (BPN/G$4$)$\text{4H}$ system. Under compressive strain (-8\%), the band structure remains nearly unchanged, resulting in minimal alteration to the current behavior. However, a notable reduction in the PVR ratio is evident, dropping to 74.84 between a bias voltage of 0.17 and 0.60 V, as presented in Table \ref{tab:pv_ratio-str}. Under tensile strain (+8\%), more pronounced effects are observed. The position of the peak region shifts significantly with respect to the bias voltage, leading to the formation of two NDR regions. The first, a weaker NDR effect of approximately 1.92, occurs between 0.06 and 0.14 V. The second, more prominent NDR region begins at 0.3 V, reaching a value of 52.39.

{\renewcommand{\arraystretch}{1.2}
\begin{table}[h!]
\caption{\label{tab:pv_ratio-str} Peak-to-valley ratio (PVR) of the electronic current ($I$) for the wave configuration in the BPN-G$_4$ system under tensile (+8\%) and compressive (-8\%) strain.}
\begin{ruledtabular}
\begin{tabular}{cccc}
strain (\%) & PVR & $V_\text{p}$ (V) & $V_\text{v}$ (V) \\
\hline
-8 & 74.84 & 0.16 & 0.60 \\
0 & 179.3 & 0.17 & 0.60 \\
+8(1$^{st}$ peak) & 1.93 & 0.06 & 0.14 \\
+8(2$^{nd}$ peak) & 52.39 & 0.32 & 0.60 \\
\end{tabular}
\end{ruledtabular}
\end{table}
}

This dual NDR behavior under tensile strain demonstrates the potential to engineer specific transport characteristics, such as the position and intensity of NDR regions, which are critical for applications in high-frequency oscillators and memory devices. Therefore, strain engineering emerges as a powerful tool to optimize and control the transport phenomena in these systems, enabling the design of more versatile and efficient nanoelectronic devices.

% \begin{figure*}
%         \includegraphics[width=\textwidth]{figs/tran/wave.pdf}
%         \caption{\label{fig:tran-4} wave}
% \end{figure*}

% \begin{figure*}
%         \includegraphics[width=\textwidth]{figs/tran/tile.pdf}
%         \caption{\label{fig:tran-5} tile}
% \end{figure*}

\section{Summary and Conclusions}

Using first-principles calculations, we have shown that functionalization by hydrogen adatoms of biphenylene/graphene heterojunctions, \bpngh, leads to the formation of catenary-like graphene ribbons, G$_\text{n}$, separated by hydrogenated BPN stripes. We found that the electronic confinement results in a metal-to-semiconductor transition, where both the width of the bandgap and its tunability by mechanical strain follow the same classification role predicted for armchair graphene nanoribbons. Finally, we have shown that the electronic current ($I$) in BPN/G presents a strong directional anisotropy, with $I$ parallel to the hydrogenated BPN stripes ($I_\text{tile}$) larger than that perpendicular to the BPN stripes ($I_\text{wave}$), where $I_\text{wave}$ presents a mechanically tunable negative differential resistance. These findings reveal the versatility \bpngh\, systems may bring important contributions to the design of electro-mechanical (nano) transducers based on carbon materials.

\begin{acknowledgments}

This work was supported by the Brazilian agencies FAPEMIG, CNPq, and CNPq - INCT (National Institute of Science and Technology on
Materials Informatics, grant n. 371610/2023-0). We would like to acknowledge computing time provided by the National Laboratory for Scientific Computing
(LNCC/MCTI) for providing HPC resources of the SDumont supercomputer.

\end{acknowledgments}	

\bibliography{name.bib}
\bibliographystyle{apsrev4-1}

\end{document}

\title{Supplementary material: Catenary-like rippled biphenylene/graphene lateral heterojunction}

{\author{Victor M. S. da C. Dias}
\affiliation{Instituto de F\'{\i}sica,  Universidade Federal de Uberlândia, Av. João Naves de Ávila, 2121 - MG, 38400-902, Brazil}

\author{Danilo de P. Kuritza}
\affiliation{Campus Avançado Jandaia do Sul, Universidade Federal do Paraná,
86900-000, Jandaia do Sul, PR, Brazil}

\author{Igor S. S. de Oliveira}
%\email{igor.oliveira@ufla.br}
\affiliation{Departamento de F\'isica, Universidade Federal de Lavras, C.P. 3037, 37203-202, Lavras, MG, Brazil}

\author{José E. Padilha}
\affiliation{Campus Avançado Jandaia do Sul, Universidade Federal do Paraná,
86900-000, Jandaia do Sul, PR, Brazil}

\author{Roberto H. Miwa}
\email{hiroki@ufu.br}
\affiliation{Instituto de F\'{\i}sica,  Universidade Federal de Uberlândia, Av. João Naves de Ávila, 2121 - MG, 38400-902, Brazil}

%\author{\textcolor{red}{Who else?}}

\date{\today}

\maketitle

\section{Phonon Spectra}

In Fig.\,SM\ref{fig:fonon_6graf}(a) and (b) we present the phonon spectra of pristine \bpng$_6$ and hydrogenate (\bpng$_6$)$_\text{4H}$. The calculations were performed by using a 2$\times$2 surface unit cell.  The absence of imaginary frequencies confirm the dynamical stability of these systems.

\begin{figure}[h!]
    \centering % Centraliza a imagem e a legenda
    \includegraphics[width=14cm]{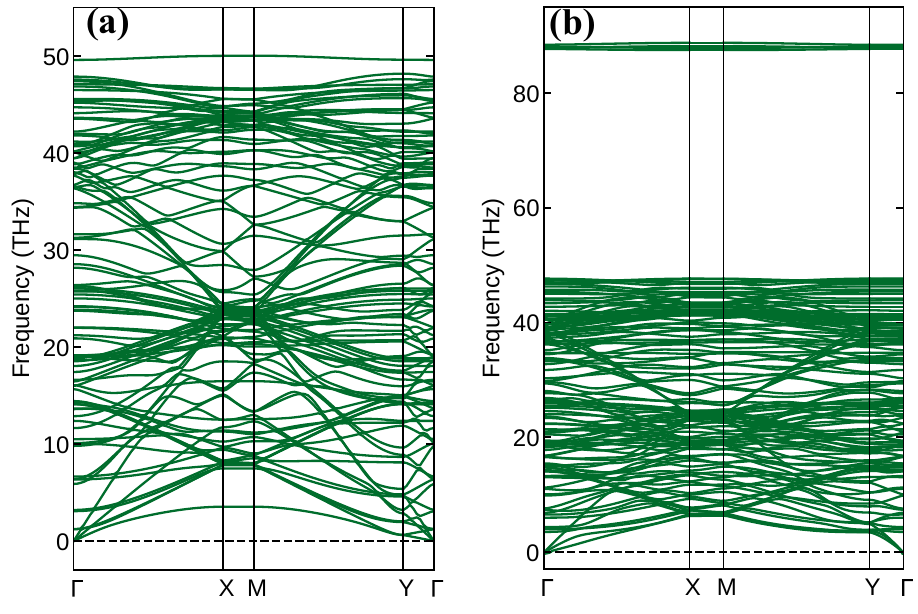}
    \caption{Phonon spectra of the \bpng$_{6}$ (a) and (BPN/G$_{6}$)$_{\text{4H}}$ (b) systems.}
    \label{fig:fonon_6graf}
\end{figure}

\newpage
\section{Electronic Band structure}

In Fig.\,SM\ref{fig:band2}(a1)-(c1) and (a2)-(c2) we present the semimetallic and semconductor electronic band structures of \bpngn, and \bpngh\, with n\,=\,6, 7, and 8, which correspond to  n\,=\,3p, 3p+1, and 3p+2 groups with p\,=\,2.

\begin{figure}[htbp]
        \includegraphics[width=\columnwidth]{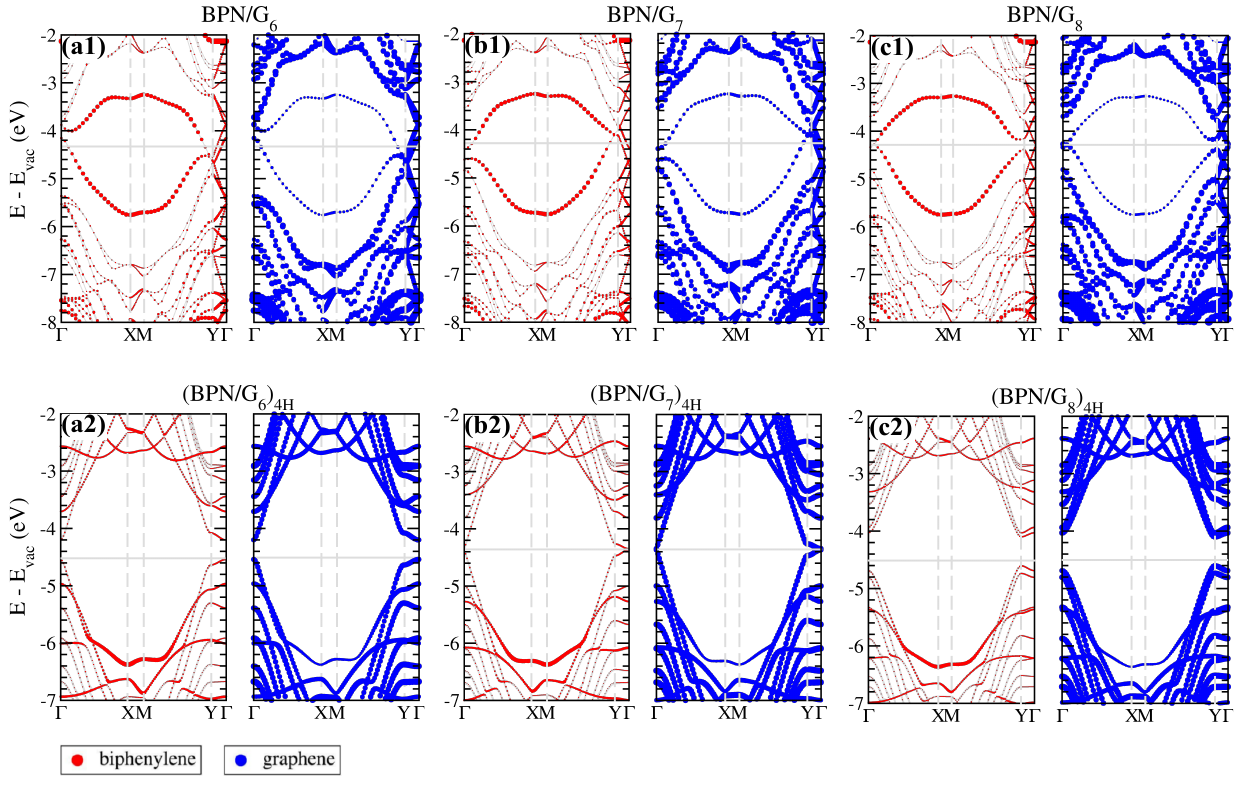}
        \caption{\label{fig:band2} Orbital projected electronic band structures of \bpngn\, and \bpngh\, with n\,=\,6\,(a1)-(a2), 7\,(b1)-(b2), and 8\,(c1)-(c2). The density of the orbital projection is proportional to the size of the filled circles. Red (blue) circles indicate the projection on the BPN (G$_\text{n}$) region.}
\end{figure}
\newpage

\section{STM IMAGE}

In Fig.\,SM\ref{fig:stm} we present the simulated STM images, empty states, of \bpngn\, with n\,=\,6, 7, and 8, where we can identify the formation of bright lines along the BNP stripes.

\begin{figure}[H]
    \centering
    \includegraphics[width= 15cm]{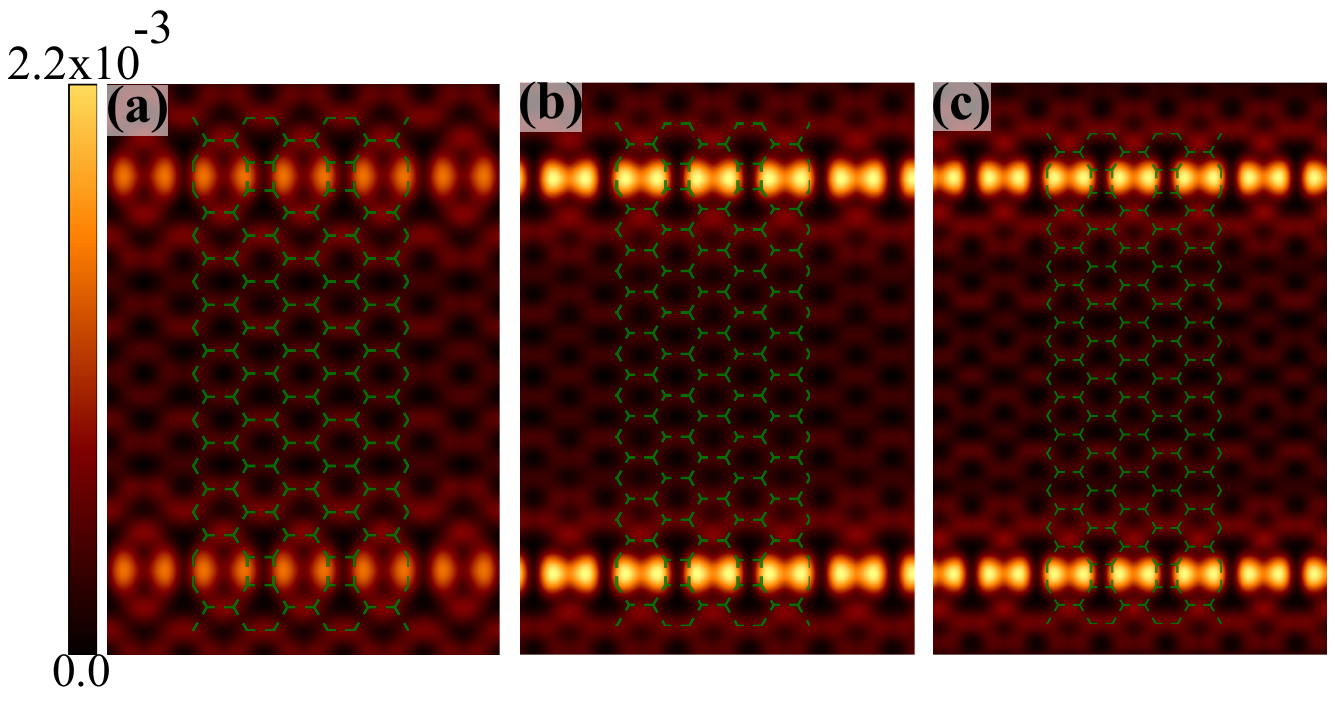}
	\caption{Simulated STM images of the  unoccupied states ($E_F+1$\,eV) of \bpngn\, with n\,=\,6\,(a), 7\,(b), and 8\,(c). The STM images are made with isovalue of [$e$/\AA$^{3}$.] }
	\label{fig:stm}
\end{figure}

\newpage
\section{Electronic transport}

In addition, the negative differential resistance (NDR) observed in the system can be attributed to the interplay between the energy-dependent transmission \( T(E) \) and the bias-induced shift in the Fermi windows of the electrodes. As shown in Fig\,SM\ref{fig:fermi-dirac}, the transmission coefficient \( T(E) \) exhibits sharp peaks at specific energies. At low bias (\(V = 0.100 \, \text{V}\)), these peaks align well with the Fermi window, defined by the difference \( f_{LE}(E) - f_{RE}(E) \), allowing significant current flow, as indicated by the integral of \( T(E) \cdot (f_{LE}(E) - f_{RE}(E)) \). For instance, at \(V = 0.175 \, \text{V}\), the product \( T(E) \cdot (f_{LE}(E) - f_{RE}(E)) \) reaches a maximum near \( E - E_F = 0.00 \, \text{eV} \).  However, as the bias increases, greater than (\(V = 0.175 \, \text{V}\), for example \(V = 0.250 \, \text{V}\)), the transmission function reduces significantly, due to the reduction of overlaps between the states in the LE and RE with the scattering region. This results in a decrease in the current contribution from these states. For example, at \(V = 0.250 \, \text{V}\), the transmission peaks are small in the Fermi window, leading to a significant reduction in the product \( T(E) \cdot (f_{LE}(E) - f_{RE}(E)) \). At higher bias (\(V = 0.300 \, \text{V}\)), the overlap becomes negligible, further suppressing the current and manifesting the NDR effect. These numerical examples clearly demonstrate the crucial role of the alignment between the transmission peaks and the Fermi-level distribution in the emergence of NDR behavior.

\begin{figure}[H]
    \includegraphics[width=\columnwidth]{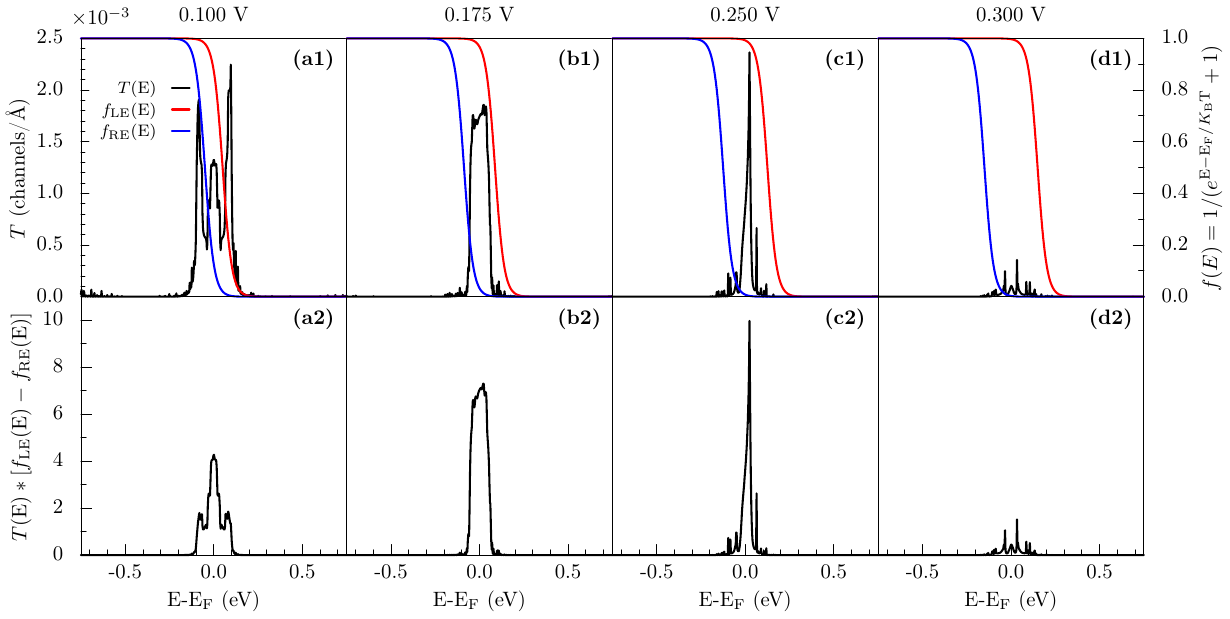}
    \caption{\label{fig:fermi-dirac} Product of the transmittance [\( T(E) \)] and the Fermi-Dirac distribution function [\( f(E) \)] for (\bpng$_4$)$_\text{4H}$, under different bias voltages.}
\end{figure}